 \documentclass[final,3p,times,twocolumn]{elsarticle}
\usepackage{amssymb}
\usepackage{graphicx}% Include figure files
\usepackage{dcolumn}% Align table columns on decimal point
\usepackage{subfigure}
\usepackage[normalem]{ulem}
\usepackage{bm}% bold math
\usepackage{fancyhdr}
\usepackage{lastpage}
\usepackage{graphicx, wrapfig, setspace, booktabs}
\usepackage[T1]{fontenc}
\usepackage[font=small, labelfont=bf]{caption}
\usepackage{fourier}
\usepackage[protrusion=true, expansion=true]{microtype}
\usepackage{hyperref}
\usepackage{url,lipsum}
\usepackage{epstopdf,epsfig}
\usepackage{amsmath,amsfonts,amssymb}
\usepackage{float}
\usepackage{url}
\usepackage{natbib}

\begin{document}

\begin{frontmatter}



%\title{Quantum-like features of an active wave-particle entity in a double-well potential}
\title{Quantum-like behavior of an active particle in a double-well potential}


%\author[inst1]{Author One}
\author[inst1,inst3]{Rahil N. Valani}%\email{rahil.valani@adelaide.edu.au}
\ead{rahil.valani@physics.ox.ac.uk}
\author[inst2]{Álvaro G. López}

%\author[inst2]{Author Two}
%\author[inst1,inst2]{Author Three}

\affiliation[inst1]{organization={School of Computer and Mathematical Sciences, University of Adelaide},
            city={Adelaide},
            postcode={5005},
            state={South Australia},
            country={Australia}}

\affiliation[inst2]{organization=  {Nonlinear Dynamics, Chaos and Complex Systems Group.\\Departamento de F\'isica},%Department and Organization
            addressline={Universidad Rey Juan Carlos, Tulip\'an s/n}, 
            city={M\'ostoles},
            postcode={28933}, 
            state={Madrid},
            country={Spain}}

\affiliation[inst3]{organization={Rudolf Peierls Centre for Theoretical Physics, Parks Road,
University of Oxford},
            city={Oxford},
            postcode={OX1 3PU},
            country={United Kingdom}}

\begin{abstract}
A macroscopic, self-propelled wave-particle entity (WPE) that emerges as a walking droplet on the surface of a vibrating liquid bath exhibits several hydrodynamic quantum analogs. We explore the rich dynamical and quantum-like features emerging in a model of an idealized one-dimensional wave-particle entity in a double-well potential. The integro-differential equation of motion for the WPE transforms to a Lorenz-like system, which we explore in detail. We observe the analog of quantized eigenstates as discrete limit cycles that arise by varying the width of the double-well potential, and also in the form of multistability with coexisting limit cycles. These states show narrow as well as wide energy level splitting. Tunneling-like behavior is also observed where the WPE erratically transitions between the two wells of the double-well potential. We rationalize this phenomena in terms of crisis-induced intermittency. Further, we discover a fractal structure in the escape time distribution of the particle from a well based on initial conditions, indicating unpredictability of this tunneling-like intermittent behavior at all scales. The chaotic intermittent dynamics lead to wave-like emergent features in the probability distribution of particle's position that show qualitative similarity with its quantum counterpart. Lastly, rich dynamical features are also observed such as a period doubling route to chaos as well as self-similar periodic islands in the chaotic parameter set.
\end{abstract}




\end{frontmatter}


\section{Introduction}

%\begin{itemize}
%    \item Introduce walking droplets from the viewpoint of hydrodynamic quantum analogs.
%    \item Highlight the features of wave-particle entity, active particle and memory in the system
%    \item Talk in more detail about the HQA especially wave-like statistics, tunneling, eigenstates that are relevant to this paper  i.e. tunneling, quantized eigenstates and wave-like statistics
%    \item Say that inspired by this, we explore the WPE in a double well potential where tunneling, wave-like statistics and quantized eigenstates may co-exist.
%    \item Mention that we use a simple Lorenz-like idealized model to explore this since we want to analyse the underlying dynamics systems and make connections between quantum-like features and features of nonlinear dynamics.
%    \item Mention what will be presented in the different sections...
%\end{itemize}

Quantization, wave-like statistics and tunneling are usually taken to be exclusive features of the microscopic quantum realm. However, in recent years, a hydrodynamic system of millimeter-sized walking droplets~\citep{Couder2005,Couder2005WalkingDroplets} has demonstrated that classical systems with memory effects can exhibit several hydrodynamic quantum analogs. In this hydrodynamic system, a millimetric droplet bounces periodically while walking horizontally on the free surface of a vertically vibrating bath of the same liquid. Each bounce of the droplet generates a localized and slowly decaying standing wave on the liquid surface. The droplet interacts with these self-generated waves on subsequent bounces resulting in horizontal motion. This gives rise to self-propelled walking and superwalking~\citep{superwalker} droplets. Three key physical properties to note about this system are: (i) The droplet and its underlying wave coexist as a \emph{wave-particle entity} (WPE); without the droplet the underlying waves decay completely. (ii) The droplet is \emph{active} in the sense of an active particle and active matter~\citep{Ramaswamy_2017}, since the droplet locally extracts energy from the vibrating liquid bath and converts it into directed motion. (iii) There is \emph{memory} in the system. Since the waves generated by the droplet decay very slowly in time, the motion of the droplet is not only influenced by the wave it generated on the most recent bounce, but also by the waves it generated in the distant past. Summarizing, the walking-droplet system constitutes a locally active system with memory effects embedded in its dynamics. This allows the particle to amplify energy fluctuations from its environment, converting them into coherent motion and breaking fundamental symmetries \cite{localact}.

Many of the hydrodynamic quantum analogs exhibited by this active WPE, in both experiments as well as simulations, typically take place in the high-memory regime and/or when the motion of the WPE is spatially confined. For example, confinement of a WPE in external potentials with a central force results in quantized orbits~\citep{Perrard2014a,Labousse_2014,labousseharmonic,Kurianskiharmonic,durey2018,Perrard2018}. By confining a WPE in a two-dimensional harmonic potential, a discrete set of orbits are observed such as circles, ovals, lemniscates and trefoils~\citep{Perrard2014a,Kurianskiharmonic}. Here, the dynamic constraint imposed on the particle by its guiding wave field results in quantum-like eigenstates emerging from wave-memory mediated interactions. Hydrodynamic analog of quantum tunneling has also been shown for a WPE. In experiments, the WPE interacting with a submerged barrier typically gets reflected from the barrier. Eddi \textit{et al.}~\citep{Eddi2009} showed that occasionally, the interactions of the WPE with the barrier can lead to droplet tunneling across the barrier. Thus, the complex interaction of the particle with its underlying wave field results in unpredictable tunneling. In the high-memory regime when the droplet is confined spatially by either applying an external potential or the presence of boundaries, one gets chaotic motion of a WPE that leads to coherent wave-like statistics~\citep{PhysRevE.88.011001,durey_milewski_wang_2020,Giletcircularcavity,Giletcircularcavity2}. These are some of the many hydrodynamic quantum analogs exhibited by the walking-droplet system (for a comprehensive review see \citep{Bush2015,Bush2020review}).

Several theoretical models have been developed over the years to capture the walker's dynamics~\citep{Turton2018,Rahman2020review}. An analytically tractable integro-differential equation of motion for the walker that captures the essential features of the horizontal walking dynamics was developed by Oza \textit{et al.}~\citep{Oza2013}. This stroboscopic model has been shown to capture several hydrodynamic quantum analogs~\citep{Oza2014,harris_bush_2014,labousse2016b,Spinstates,Kurianskiharmonic,Tambascoorbit,ValaniHOM,Saenz2021}. % and also results in rich dynamical behaviors for walkers~\citep{ValaniUnsteady,Durey2020lorenz,twodroplets,Oza2014a,Tambasco2016,PhysRevFluids.3.013604,PhysRevFluids.2.053601}. 
 To explore the dynamics of WPEs and their hydrodynamic quantum analogs beyond the restricted parameter space of experiments, the framework of generalized pilot-wave dynamics was developed~\citep{Bush2015}. It is a theoretical abstraction rooted in the walking-droplet system that allows for exploration of a broader class of dynamical systems and the discovery of new quantum analogs~\citep{Bush2020review}. The generalized pilot-wave framework has motivated exploration of idealized pilot-wave systems that consider dynamics of WPEs in one horizontal dimension where the integro-differential equation of motion reduces to Lorenz-like dynamical systems~\citep{phdthesismolacek,Durey2020,Durey2020lorenz,ValaniUnsteady,Valanilorenz2022}. Such Lorenz-like models of one-dimensional WPEs have shown rich features that emerge for a WPE in different scenarios such as anomalous transport of WPE in a linear potential~\citep{ValaniHOM} and interference effects for a WPE in a sinusoidal potential~\citep{Perks2023}.

By using a simple Lorenz-like model, in this paper we explore hydrodynamic quantum analogs of a one-dimensional WPE in a double-well potential. A quantum particle in a double-well potential is a canonical setup modeling many physical situations such as hydrogen bonding~\citep{DWexample1}, proton transfer in DNA~\citep{DWexample2} and quantum logic gates~\citep{DWexample3}. By considering an analogous setup of a one-dimensional WPE in an external double-well potential, we explore the resulting rich dynamics and emergent quantum-like features. 

The paper is organized as follows. In Sec.~\ref{Sec: DE} we present the equation of motion for the WPE in a double-well potential and show its transformation to a Lorenz-like dynamical system. We then explore stationary states of the WPE and their linear stability in Sec.~\ref{Sec: LS}. In Sec.~\ref{Sec: Quantization} we describe analog of quantized eigenstates, followed by analogs of quantum tunneling and wave-like statistics in Secs.~\ref{Sec: tunneling} and \ref{Sec: wavelike stats}, respectively. We further study the organization of these dynamical features of our system in the parameter space in Sec.~\ref{Sec: wavelike stats}. We discuss and conclude in Sec.~\ref{Sec: Discussion and Conclusion}.

\begin{figure}
\centering
\includegraphics[width=\columnwidth]{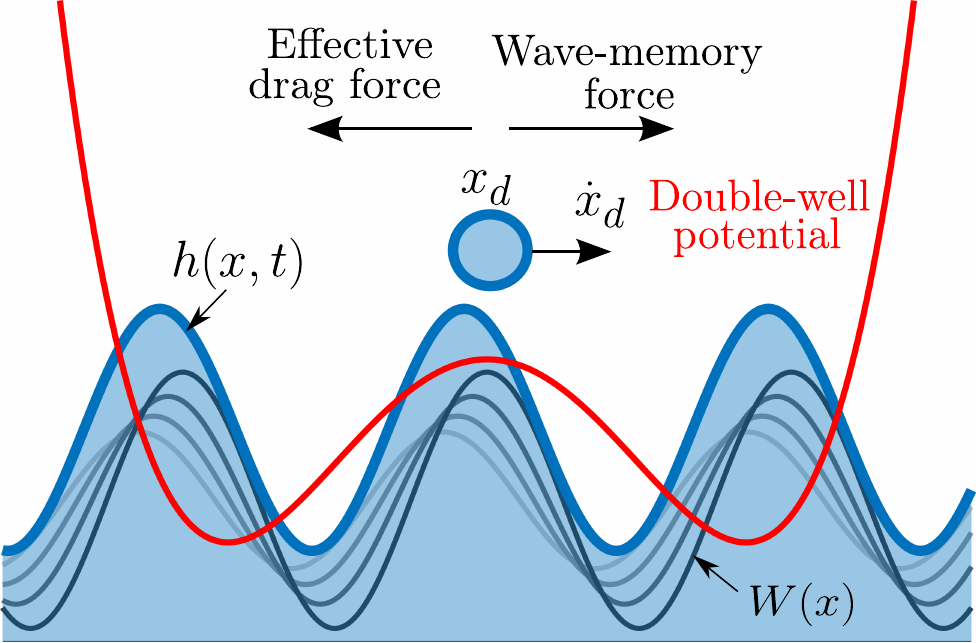}
\caption{Schematic of the system (in dimensionless units) showing a one-dimensional active WPE (blue) in a symmetric double-well potential (red). A particle of unit dimensionless mass located at $x_d$ and moving horizontally with velocity $\dot{x}_d$ experiences a wave-memory force from its self-generated wave field (blue filled area), and an effective drag force. The underlying wave field is a superposition of the individual waves of spatial form $W(x)=\cos(x)$, which are generated by the particle continuously along its trajectory and decay exponentially in time (black and gray curves). An external double-well potential exerts an additional nonlinear force $F(x_d)=-Ax_d^3+B x_d$ on the particle.}
\label{Fig: schematic}
\end{figure}

\begin{figure*}
\centering
\includegraphics[width=2.0\columnwidth]{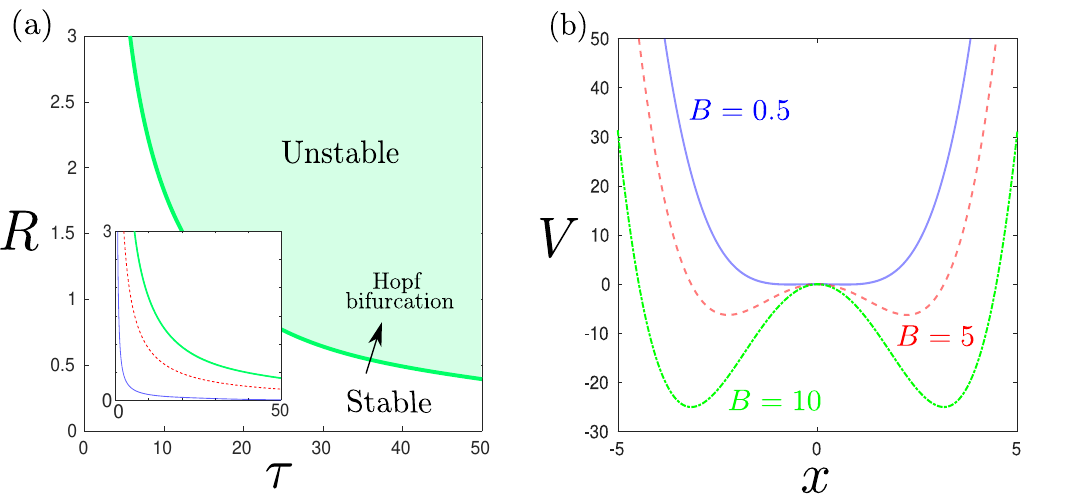}
\caption{Linear stability analysis of a stationary WPE at minima in a double-well potential. (a) Linear stability curves (see Eq.~\eqref{eq: lin stab}) for the stationary states of WPE in minima of a (b) double-well potential $V(x)=Ax^4/4-B x^2/2$ for a fixed $A=1$ and $B=0.5$ (blue solid), $B=5$ (red dashed) and $B=10$ (green dashed-dotted). In panel (a), below the stability curve (green), the stationary state is stable to small perturbations. As the curve is crossed, a Hopf bifurcation occurs, and the stationary state becomes unstable to small perturbations in an oscillatory manner. The inset shows how the stability curve changes for different values of $B$ as stated above. The width of the potential (horizontal distance between the minima) scales as $\sqrt{B}$ while the height of the potential (vertical distance between the maximum and minima) scales as $B^2$.}
\label{Fig: lin stab}
\end{figure*}

\section{Dynamical equations}\label{Sec: DE}

Consider a particle (droplet) located at position $x_d$ and moving horizontally with velocity $\dot{x}_d$, while continuously generating standing waves with prescribed spatial structure $W(x)$, which decay exponentially in time. The particle is subjected to an external symmetric double-well potential of the form $V(x)=\bar{A} x^4/4 - \bar{B} x^2/2$. Oza \textit{et al.}~\citep{Oza2013} developed a theoretical stroboscopic model that averages over the vertical periodic bouncing motion of the particle and provides a trajectory equation for the horizontal walking dynamics of a WPE in $2$D. A reduction of this model to describe the dynamics of a $1$D WPE is given by the following integro-differential equation of motion~\citep{Oza2013,Durey2020,Durey2020lorenz,ValaniUnsteady} 
\begin{align}\label{Eq: dimensional eq}
    &m\ddot{x}_d+D\dot{x}_d = \\ \nonumber
    &\frac{F_m}{T_F}\int_{-\infty}^t f(k_F(x_d(t)-x_d(s)))\,\text{e}^{-(t-s)/T_F \text{Me}}\,\text{d}s + F(x_d).\\ \nonumber 
\end{align}
The left-hand-side of Eq.~\eqref{Eq: dimensional eq} is composed of an inertial term $m \ddot{x}_d$ and an effective drag term $D \dot{x}_d$, where the overdot denotes a time derivative. The first term on the right-hand-side of the equation captures the time-delayed self-force on the droplet mediated by its underlying wave field. This non-conservative force is proportional to the gradient of the underlying wave field. The wave field is calculated through integration of the individual wave forms $W(x)$ that are continuously generated by the particle along its trajectory, and decay exponentially in time. The function $f(x)=-W'(x)$ is the negative gradient of the wave form created by the particle. The integration of past wave forms produces a feedback between the droplet and the underlying fluid, yielding a differential equation with distributed time-delay. The second term on the right-hand-side is the conservative force on the particle from the double-well potential given by 
$$F(x_d)=-\frac{\text{d}V(x)}{\text{d}x} \Big{|}_{x=x_d}=-\bar{A}x_d^3+\bar{B} x_d.$$
The parameters in Eq.~\eqref{Eq: dimensional eq} are as follows: $m$ is the droplet mass, $D$ is the effective time-averaged drag coefficient, $k_F=2\pi/\lambda_F$ is the Faraday wavenumber with $\lambda_F$ the Faraday wavelength, $F_m=m g A_m k_F$ is a non-negative wave-memory force coefficient with $g$ as the gravitational acceleration and $A_m$ the amplitude of surface waves, $\text{Me}$ is the memory parameter that describes the proximity to the Faraday instability and $T_F$ is the Faraday period, i.e. the period of droplet-generated standing waves and also the bouncing period of the walking droplet. We refer the interested reader to the literature \cite{Oza2013} for more details and explicit expressions for these parameters. 

If Eq.~\eqref{Eq: dimensional eq} is non-dimensionalized using $t'=D{t}/m$ and $x'=k_F x$, and dropping the primes on the dimensionless variables, we get
\begin{align}\label{eq: dimless eq2}
    \ddot{{x}}_d +  \dot{{x}}_d =  R \int_{-\infty}^{t} f\left( {x}_d(t)-{x}_d(s) \right)\,\text{e}^{-\frac{(t-s)}{{\tau}}}\,\text{d}s-{A}x^3_d + {B}x_d,
\end{align}
where we have introduced a dimensionless wave-amplitude $R= m^3 g A_m k_F^2/(D^3 T_F)$ and a dimensionless decay time of waves or the memory of the system $\tau=D T_F \text{Me}/m$. These two parameters are the key parameters introduced to study the effects of the self-force on the
quantum-like features of the self-propelled particle. Two more dimensionless parameters represent tacitly the shape (depth and width) of the external double-well potential: $A=m\bar{A}/(D^2 k^2_F)$ and $B=m\bar{B}/D^2$. The theoretical setup is schematically shown in Fig.~\ref{Fig: schematic}.

The experimental wave-form $W(x)$ of walking and superwalking droplets has features of spatial oscillations as well as spatial decay~\citep{Molacek2013DropsTheory,superwalkernumerical}. By neglecting spatial decay and only including spatial oscillations, we choose a simple sinusoidal particle-generated wave form such that $W(x)=\cos(x)$ and $f(x)=\sin(x)$.This simple choice still captures the key dynamical features of a WPE~\citep{Durey2020hqft,Durey2020lorenz,ValaniUnsteady} and it allows us to transform the integro-differential equation of motion in Eq.~\eqref{eq: dimless eq2} into a simplified Lorenz-like system of ordinary differential equations (ODEs). In this manner, an infinite-dimensional dynamical system in the phase space, is reduced to a much more manageable four-dimensional system. This simplified system might also be relevant to similar problems in electrodynamics of extended bodies \citep{onanelec}, where viscous effects are not present and time-delayed self-forces can also produce nonlinear self-oscillations \citep{jenkins} via Hopf bifurcation. 

The transformation gives us (see \cite{Valanilorenz2022,Durey2020lorenz} for details) the nonlinear four-dimensional ODE
\begin{equation}
    \label{lorenz}
    \begin{split}
    \dot{x}_d&=X, \\
    \dot{X}&=Y-X-A x^3_d + B x_d, \\
    \dot{Y}&=-\frac{1}{\tau}Y+XZ, \\
    \dot{Z}&=R-\frac{1}{\tau}Z-XY.
  \end{split}
\end{equation}
Here, $X=\dot{x}_d$ is the particle velocity, $$Y=R \int_{-\infty}^{t} \sin(x_d(t)-x_d(s))\,\text{e}^{-\frac{(t-s)}{{\tau}}}\,\text{d}s$$ is the wave-memory force on the particle (proportional to the wave gradient at the particle location) and $$Z=R\int_{-\infty}^{t} \cos(x_d(t)-x_d(s))\,\text{e}^{-\frac{(t-s)}{{\tau}}}\,\text{d}s$$ is the wave field height at the particle location. For the remainder of this paper, we will analyze in detail the Lorenz-like dynamical system in Eq.~\eqref{lorenz}. The simulations presented in this paper were performed by numerically solving Eq.~\eqref{lorenz} in MATLAB using the inbuilt solver ode45.

\begin{figure*}
\centering
\includegraphics[width=2\columnwidth]{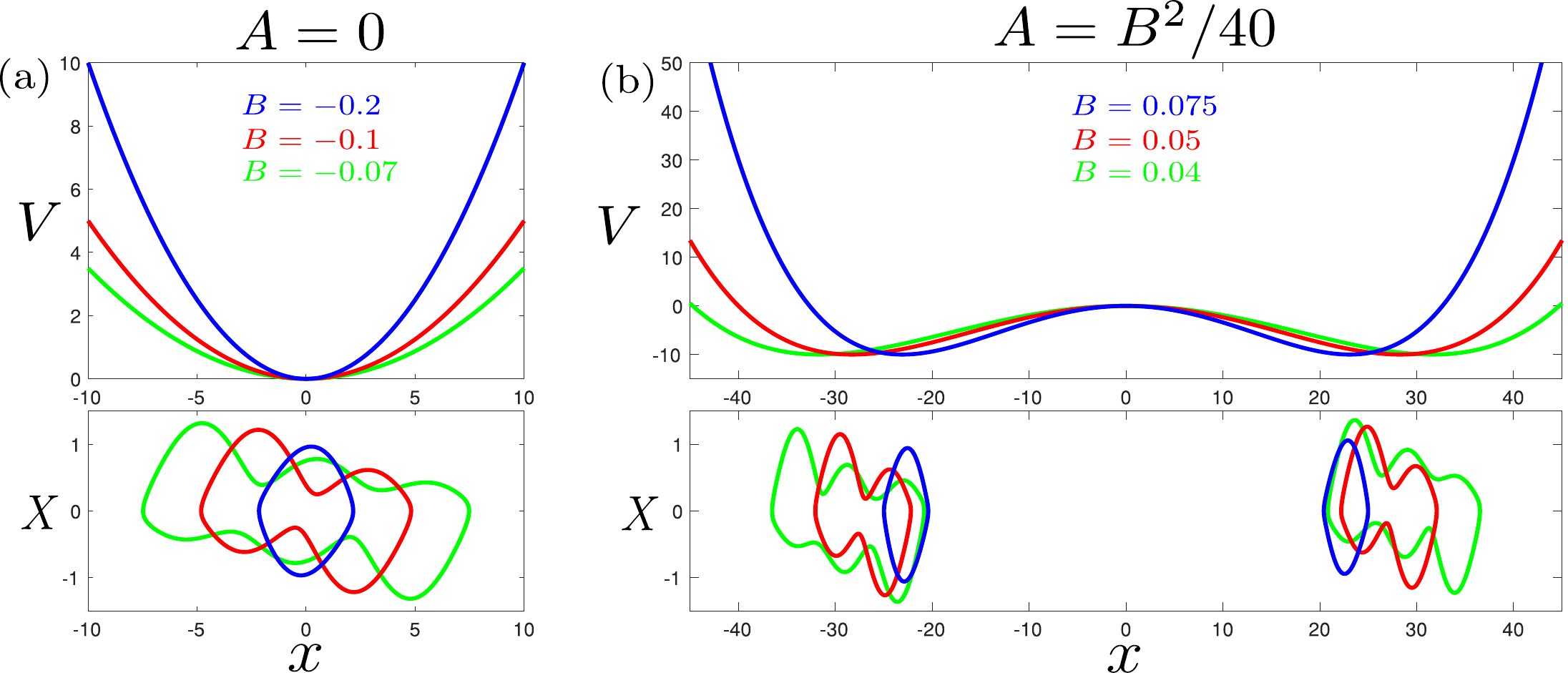}
\caption{Limit cycles as analog of quantized eigenstates for WPE in (a) harmonic potential and (b) double-well potential as the width of the potential is varied. For (a) and (b), the top panel shows the external potential while the bottom panel shows the phase-space trajectories in the $(x_d,X)$ phase-plane. The parameters for the harmonic potential in panel (a) are fixed $A=0$ and varying $B=-0.2$ (blue), $-0.1$ (red) and $-0.07$ (green), while the parameters for the double-well potential in panel (b) are fixed $H=10$ and varying $A$ and $B$ with $B=0.075$ (blue), $0.05$ (red) and $0.04$ (green) and corresponding $A=B^2/(4H)$. The WPE parameters are fixed to $R=0.5$ and $\tau=3$.}
\label{Fig: eigenstates}
\end{figure*}

\section{Stability analysis}\label{Sec: LS}

By making the time derivatives to zero in Eq.~\eqref{lorenz}, we can find the stationary states of a WPE in the double-well potential. This results in the following particle equilibria
$$(x_d,X,Y,Z)=(0,0,0,R\tau),$$
$$(x_d,X,Y,Z)=(\pm\sqrt{B/A},0,0,R\tau).$$
These particle equilibria correspond to a stationary WPE located at maximum ($x=0$) and the two symmetric minima ($x=\pm\sqrt{B/A}$) of the double-well potential, respectively. To understand the stability of these two equilibrium points, which correspond to stationary states of the WPE, we perform a linear stability analysis of these equilibrium states.

\subsection{Linear stability of a stationary WPE at maximum of the double-well}

Performing a linear stability analysis~\citep{strogatz} by applying a small perturbation to a stationary WPE located at maximum of the double-well, $(x_d,X,Y,Z)=(0,0,0,R\tau)+\epsilon(x_{d1},X_1,Y_1,Z_1)$, where $\epsilon>0$ is a small perturbation parameter, one gets the following characteristic polynomial for the eigenvalues $\lambda$ that determine the growth rate of perturbations:
\begin{equation*}
    \left(\lambda+\frac{1}{\tau}\right)\left( \lambda^3 + \left( 1+\frac{1}{\tau} \right)\lambda^2 + \left(\frac{1}{\tau}-R\tau-B\right)\lambda -\frac{B}{\tau} \right) = 0.
\end{equation*}
Apart from the trivial eigenvalue $\lambda=-1/\tau$, three more eigenvalues can be found by solving the cubic equation in the second parenthesis. Since the parameters $R,\tau,B>0$, using Descartes rule of sign, we find that there is always at least one positive real root of this equation making this stationary state unstable to small perturbations. Simply put, the addition of the wave-particle self-force cannot stabilize the particle equilibrium the center of the double-well potential.

\subsection{Linear stability of a stationary WPE at minima of the double-well}

By applying a small perturbation to a stationary WPE located at minima of the well, $(x_d,X,Y,Z)=(\pm\sqrt{B/A},0,0,R\tau)+\epsilon(x_{d1},X_1,Y_1,Z_1)$, where $\epsilon>0$ is a small perturbation parameter, we get the following characteristic equation for the eigenvalues $\lambda$:
\begin{equation}
    \left(\lambda+\frac{1}{\tau}\right)\left( \lambda^3 + \left( 1+\frac{1}{\tau} \right)\lambda^2 + \left(\frac{1}{\tau}-R\tau+2B\right)\lambda +\frac{2B}{\tau} \right) = 0.
    \label{eq:polynom2}
\end{equation}
Again, the non-trivial eigenvalues can be found by solving the cubic equation in the second parenthesis.
%\begin{equation}\label{eq: cubic}
%\lambda^3 + \left( 1+\frac{1}{\tau} \right)\lambda^2 + \left(\frac{1}{\tau}-%\end{equation}
Using Descartes rule of sign we find that: (i) if $2B\tau+1-R\tau^2>0$, then there are either three negative real roots or one negative real root and a complex conjugate pair, and (ii) if $2B\tau+1-R\tau^2<0$, then there are either one negative and two positive real roots, or one negative real root and a complex conjugate pair. Hence, the stability of this stationary state is determined by real part of the complex conjugate pair, whenever they exist. By substituting $\lambda=i\omega$, one can separate the real and imaginary parts of the third order polynomial appearing in Eq.~\eqref{eq:polynom2} and find the stability boundary where this complex conjugate pair changes the sign of its real part. This gives the following stability boundary in the $(\tau,R)$ parameter space (a hypersurface in the entire four-dimensional parameter space), which only depends on  the parameter $B$ of the external potential:
\begin{equation}\label{eq: lin stab}
    R=\frac{1}{\tau^2}+\frac{2B}{\tau+1}.
\end{equation}
This critical boundary gives us the values of the Hopf bifurcation, above which the rest state becomes unstable and fluctuations in the particle's dynamics take place, yielding quantized orbits in terms of limit cycles.
From a physical point of view, above this boundary in the parameter space, the stationary WPE at the minima of the double-well becomes unstable in an oscillatory sense. Importantly, the frequency of the small amplitude oscillations right above the critical boundary is independent of the external potential, and given by
\begin{equation*}
    \omega = \sqrt{R-\frac{1}{\tau^2}},
\end{equation*}
which is always real, since just above the stability boundary, one gets $R>1/\tau^2$~(see Eq.~\ref{eq: lin stab}). This frequency of horizontal motion is different from the frequency of the droplet's bouncing on the fluid. This transformation of a vertical bath vibrations into a horizontal oscillation with non-resonant frequencies via Hopf bifurcation constitutes a hallmark of self-oscillation \cite{jenkins}. Figure~\ref{Fig: lin stab} shows the stability boundary in the parameter space $(\tau,R)$ for different values of $B$ and the corresponding double-well potentials. Using the parameters of the double-well potential $A$ and $B$, we can define two new parameters $W$ and $H$ corresponding to the width and the height of the potential, respectively. We define the width of the double-well potential as the horizontal distance between the two minima giving $W=\sqrt{4B/A}$, and height as the vertical distance between the minima and the local maximum giving $H=B^2/(4A)$. In this parameter representation, the potential can be written as $V(x)=16 H ((x/W)^4-(x/W)^2)$, acquiring its most explicit geometrical interpretation. Since $W \propto \sqrt{B}$ and $H \propto B^2$, we see in Fig.~\ref{Fig: lin stab}(b), as $B$ is increased for fixed $A$, the width and height of the double-well potential increase. This results in an upward shift in the stability boundary for stationary states as $B$ increases (see Fig.~\ref{Fig: lin stab}(a)) indicating that larger wave-amplitude and memory are required to make the stationary state unstable.

\section{Multistability of quantized states and energy splitting}\label{Sec: Quantization}

Once the stationary states of a WPE in the double-well potential become unstable, limit cycle oscillations emerge in the low-memory regime (i.e. small $\tau$ above the instability). In this section, we explore the nature of these limit cycle oscillations, specifically in relation to the analog of quantized eigenstates.

\subsection{Quantized states}

For a classical point particle in a double-well potential (i.e. without dissipation and self-forces in Eq.~\ref{lorenz}), the equations of motion are given by
\begin{equation*}
    \label{classical}
    \begin{split}
    \dot{x}_d&=X, \\
    \dot{X}&=-A x^3_d + B x_d. \\
  \end{split}
\end{equation*}
This takes the form of a Hamiltonian dynamical system with total energy being a constant of motion. Hence, the system dynamics result in a continuous spectrum of periodic orbits for the particle that correspond to a continuous spectrum of its energy~\citep{strogatz}. Conversely, for a quantum particle in a double-well potential, solving the Schr\"odinger equation gives a countably infinite number of discrete eigenstates with corresponding discrete energy spectrum~\citep{doublewell1}. 

The dynamical system for our WPE in a double-well potential is non-conservative due to the presence of dissipative forces as well as self-forces from the underlying wave field. Perrard \textit{et al.}~\citep{Perrard2014a} experimentally showed that walking droplets subjected to an external harmonic potential in two dimensions can self-organize in various ``quantized eigenstates" corresponding to periodic orbits (i.e. limit cycles of the dynamical system) such as circles, ovals, lemniscates and trefoils. These different shapes of limit cycles emerge as the width of the external harmonic potential is increased. Using their sophisticated theoretical model of walking droplets, Durey \textit{et al.}~\citep{durey2018} showed similar quantized eigenstates for a walker confined to move in one horizontal dimension under the influence of an external harmonic potential. Thus, a hydrodynamic analog of quantized eigenstates was formed for the walking droplet where different geometries of limit cycles (with quantized amplitude and/or angular momentum) came into existence as the width of the harmonic potential was varied. 

Our simple Lorenz-like system also captures this analog of quantized eigenstates. By choosing $A=0$ and $B<0$, our double-well potential transforms to a simple harmonic potential and we obtain limit cycles with different number of ``lobes" as the width of the harmonic potential is increased resulting in quantized amplitudes. The first three limit cycles in the $(x_d,X)$ phase-plane along with the harmonic potential are shown in Fig.~\ref{Fig: eigenstates}(a).

Now, extending this idea to the double-well potential, we obtain similar quantized states in each of the minima by keeping the height $H$ fixed and varying the width $W$ of the potential. We observe quantized limit cycle states similar to the harmonic potential emerging in each of the minima of the double-well potential as shown in Fig.~\ref{Fig: eigenstates}(b). 

\begin{figure*}
\centering
\includegraphics[width=2\columnwidth]{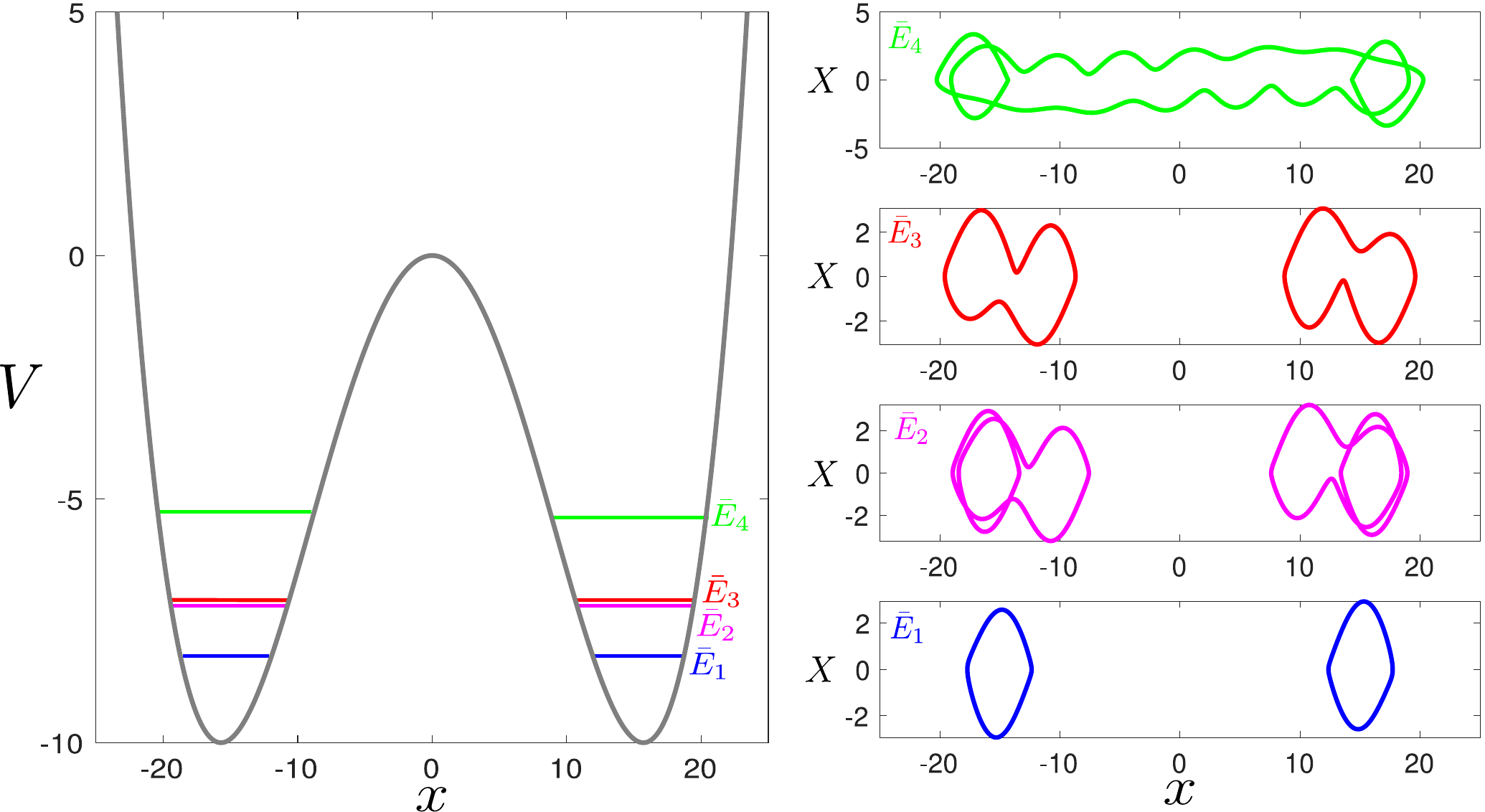}
\caption{Multistable quantized states and energy splitting for a WPE in a double-well potential. Left panel shows the double-well potential with average Lyapunov energies as colored horizontal lines corresponding to distinct limit cycle states shown on the right panels. The different limit cycles coexist for the same parameter values but different initial conditions. Parameter values were fixed to $R=3$, $\tau=1.9$, $H=10$, $W=5 \times 2\pi$, $B=16 H/W^2$ and $A=64 H/W^4$.}
\label{Fig: eigen multi}
\end{figure*}

\begin{figure}
\centering
\includegraphics[width=\columnwidth]{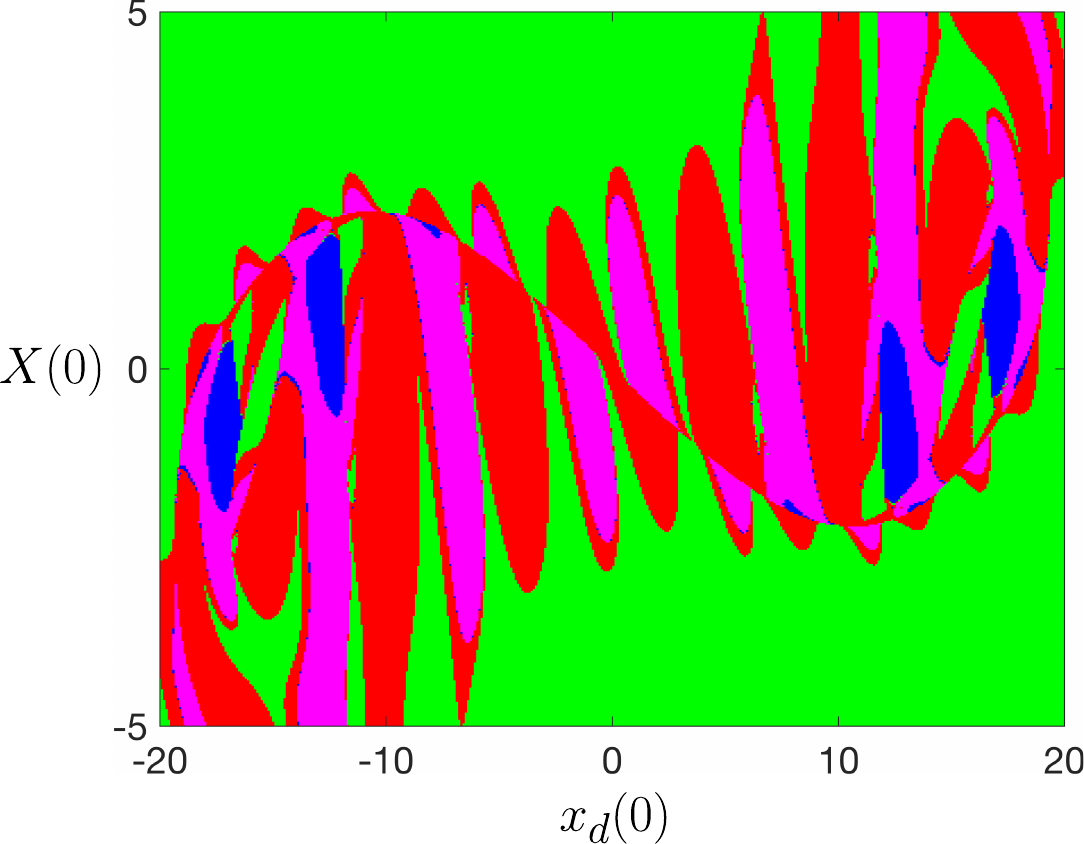}
\caption{Basin of attraction in the initial condition plane $(x_d(0),X(0))$ keeping $Y(0)=Z(0)=0$ fixed. The parameters and colors correspond to limit cycle states in Fig.~\ref{Fig: eigen multi} with blue, magenta, red and green representing states with increasing Lyapunov energies $\bar{E}_1, \bar{E}_2, \bar{E}_3$ and $\bar{E}_4$ respectively. The resolution of this plot is $400$ by $400$.}
\label{Fig: eigen basin}
\end{figure}

\begin{figure*}
\centering
\includegraphics[width=1.75\columnwidth]{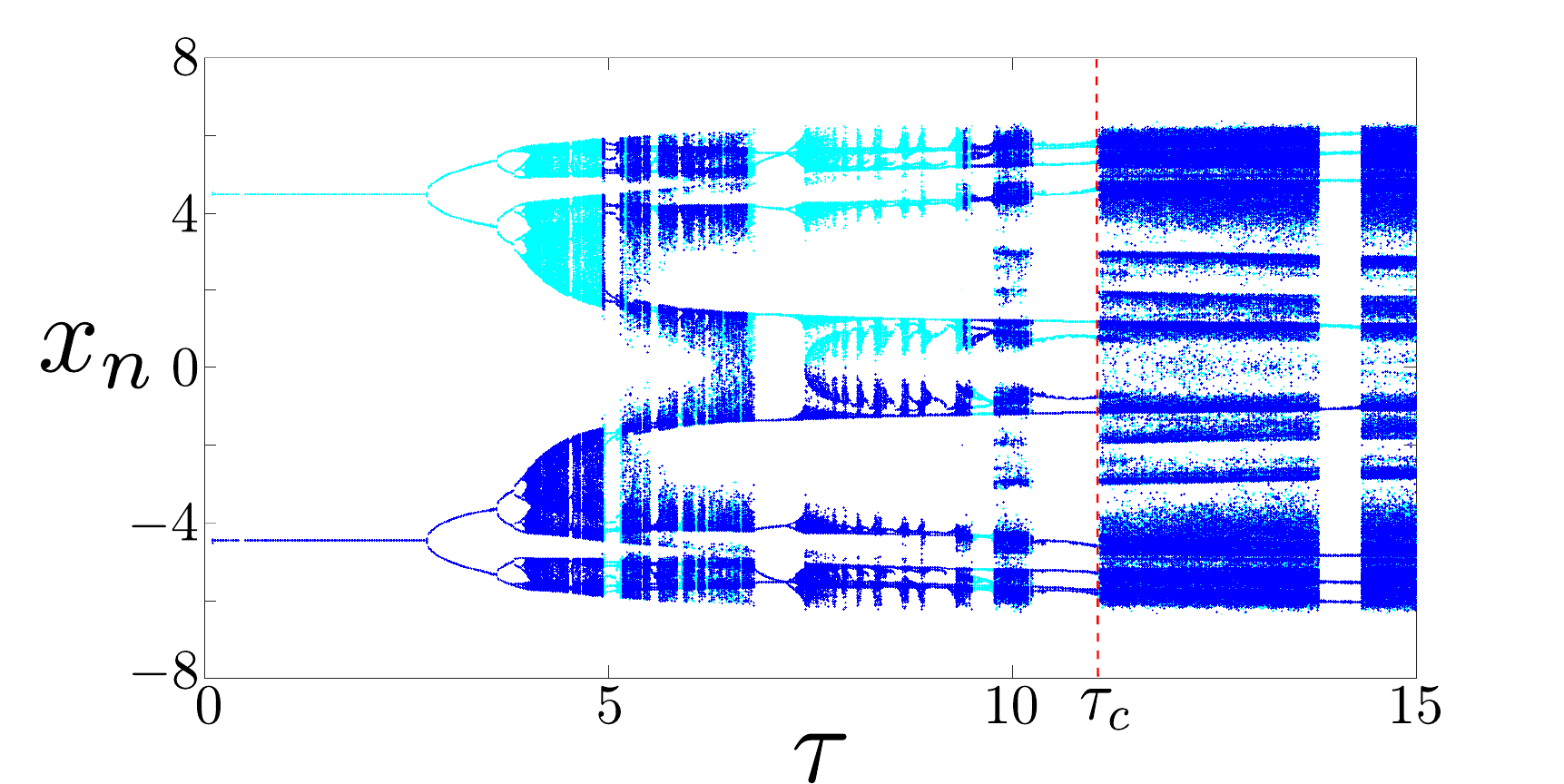}
\caption{Bifurcation diagram of a WPE in a double-well potential as a function of the memory parameter $\tau$. The values $x_n$, particle position when the instantaneous velocity is zero, are plotted as a function of $\tau$. Blue and cyan indicate two different but symmetric initial conditions used in simulations for each value of $\tau$. The red dashed line indicates the onset of crises-induced intermittency at the critical memory of $\tau_c\approx11.20175$. Other parameters are fixed to $R=1.2$, $B=2$, $H=10$ and $A=B^2/4H$.}
\label{Fig: bif}
\end{figure*}

\subsection{Multistability and energy splitting}\label{Sec: Multistability}
 The above analog of quantized orbits that has been demonstrated for walking droplets based on varying the width of the harmonic potential is not entirely satisfactory, since one would like the existence of all quantized states for a fixed external potential i.e. without changing the parameters of the dynamical system. This notion exists in dynamical systems and is known as multistability, when different attractors (e.g. fixed points, limit cycles or strange attractors) coexist for the same parameter values but different initial conditions. For a harmonic potential, multistability with two or three limit cycles is evident in the discrete-time model of Durey \textit{et al.}~\cite{durey2018} and it is also realized in the Lorenz-like model Eq.~\eqref{lorenz} considered in this paper for $A=0$. We now turn to examine multistability and coexistence of different energy levels inside the external double-well potential, which yields a finite spectrum of quantized states at different energy levels. 

For our WPE in a double-well potential, we find the coexistence of different limit cycles for the same parameter value, with their realization being dependent on the initial conditions. An example is shown in Fig.~\ref{Fig: eigen multi}, where four different types of ``quantized eigenstates" coexist: (i) limit cycles with one lobe (blue) where the WPE is confined either in the left or right minimum of the double well potential, (ii, iii) two slightly different limit cycles both having a two-lobe structure (magenta and red) and existing near each minima, and (iv) a limit cycle encompassing both the minima of the double well (green). For each of these states, we assign an energy given by the Lyapunov energy function 
$$E(x_d(t))=\frac{1}{2}\dot{x}_d(t)^2 + V(x_d(t)).$$
This function has traditionally been used to study the global stability of dissipative dynamical systems \cite{yorke}. Our Lyapunov energy function contains the conservative part of the energy associated with our system and assigns a mechanical energy to the oscillator. The active particle in the double-well potential can also be thought of as a self-oscillator~\cite{jenkins} where the self-propulsion and dissipation mechanisms draw and release energy, respectively, so that the average of the Lyapunov energy function is conserved along a limit cycle. Such self-oscillators behave like thermodynamic engines and are thus endowed with a thermodynamic efficiency~\cite{therlor}. %The dissipation and the self-propulsion alternatively release and draw energy into the system, so that the average of the Lyapunov energy function is conserved along the limit cycle.

For a given limit cycle state, the mean Lyapunov energy $\bar{E}$ can be calculated by computing the time average of the Lyapunov energy function along the limit cycle (colored horizontal lines in left panel of Fig.~\ref{Fig: eigen multi}). We find that the value of the mean Lyapunov energy is well separated for blue, magenta/red and green states. Thus, we have coexisting distinct ``eigenstates" as limit cycles with splitting of energy in terms of the mean Lyapunov energy function. We note that emergence of multistable orbits in phase-space as an analog of quantization was also shown in a state-dependent time-delay system~\citep{LOPEZ2023113412},  inspired by self-interactions of moving electrodynamic bodies. In our wave-particle system too, the wave-memory force resulting from self-interactions provides the physical mechanism for the quantization of orbits.

We further note that the mean Lyapunov energy values of the magenta and red states are very closely spaced. This is a characteristic feature that emerges when solving the Schr\"odinger equation for a quantum particle inside a double-well potential~\citep{doublewell1,doublewell2,doublewell3}. These closely spaced energy levels are due to similar looking eigenstates for the quantum particle. For the quantum system, this can be understood in terms of the finite height of the central maximum of the double-well potential. If the central barrier of the double-well was infinitely high, then the energy spectrum of the total system would be doubly degenerate. The finite height of the central barrier lifts the degeneracy and results in the splitting of eigenstates, with closely spaced energies. For our WPE system, we also observe that the two limit cycles states (magenta and red) are only slightly different geometrically and dynamically in phase space. Hence, we get a very small difference for the mean Lyapunov energies for these two states forming an analog of degenerate level splitting. %Hence, the hyperfine splitting in our Lorenz-like dynamical system can be thought of as an analog to hyperfine splitting of a quantum particle in a double well potential.

To further understand the coexistence of these different limit cycles as quantized eigenstates for the WPE, we plot the basin of attraction for each state in the initial condition space $(x_d(0),X(0))$ while keeping fixed $Y(0)=0$ and $Z(0)=0$. Recall, the basin of attraction is a plot that assigns a definite color to each initial condition in some region of the phase space, depending on its asymptotic behavior. The corresponding structure is shown in Fig.~\ref{Fig: eigen basin}. We see that the basin of different limit cycle states exhibit intricate structure. However, despite the existence of self-similarity for the limit cycle with energy $\bar{E}_1$ with respect to the pair with energies $\bar{E}_2$ and $\bar{E}_3$, they are not intermingled, nor they present the Wada property. Therefore, the transients are not specially interesting from the point of view of transient chaos in the present model.%, as compared to previous studies, where the role of transients in self-excited systems have been stressed \cite{LOPEZ2023113412}.

Randomly starting initial conditions can result in any of these limit cycle states with probability given by the fraction of the area occupied by each of these states in the basin of attractions. This concept is formally known as basin stability \citep{bastab}, and these probabilities depend on the domain of the phase space inspected. For example, for the phase space domain shown in Fig.~\ref{Fig: eigen basin}, we can calculate that the probability of obtaining the green state is $50.82\%$, the red state is $29.74\%$, the magenta state is $16.20\%$ and the blue state is $3.24\%$. These percentages provide an approximate picture of the degree of stability of each limit cycle in this particular domain. 
%Systematic numerical computations inside the region indicate some degree of self-similarity in the basins, although fractality is rather weak in the present system, as compared to previous studies ~\citep{LOPEZ2023113412} with time-delayed dynamical systems. 
%The basins are smooth and, despite the existence of self-similarity for the limit cycle $E_1$ with respect to the the pair $E_2$ and $E_3$, they are not intermingled, nor exhibit the Wada property.

\begin{figure*}[t]
\centering
\includegraphics[width=2\columnwidth]{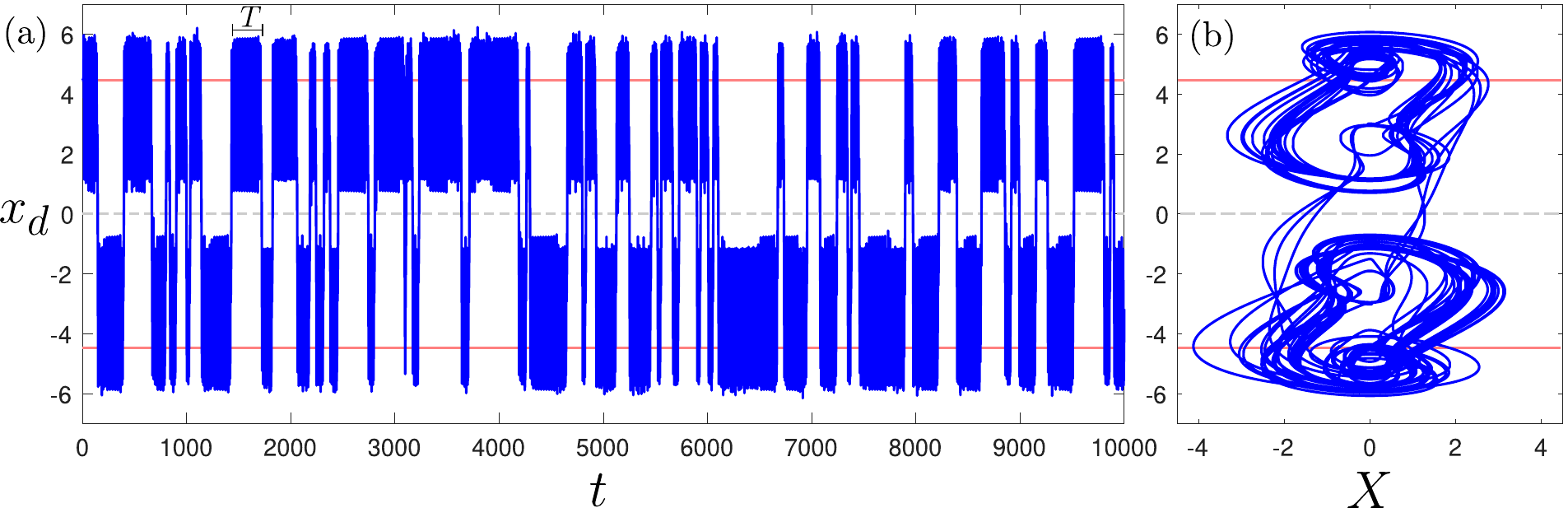}
\caption{Tunneling-like behavior via intermittent dynamics of a WPE in a double-well potential. (a) Space-time trajectory of the particle undergoing intermittent motion between the two wells of the double well potential with residence time $T$. Solid line indicates the minima of the potential while dashed line indicates the maximum. See Supplemental Video S1 for a movie. Here, $\tau=11.21$, $R=1.2$, $B=2$, $H=10$ and $A=B^2/4H$. (b) Projection of the phase-space trajectory in $(X,x_d)$ plane showing the intermittent motion as erratic jumping between two chaotic attractors - one corresponding to each minima of the double well.}
\label{Fig: intermittency}
\end{figure*}

\begin{figure*}[t]
\centering
\includegraphics[width=2\columnwidth]{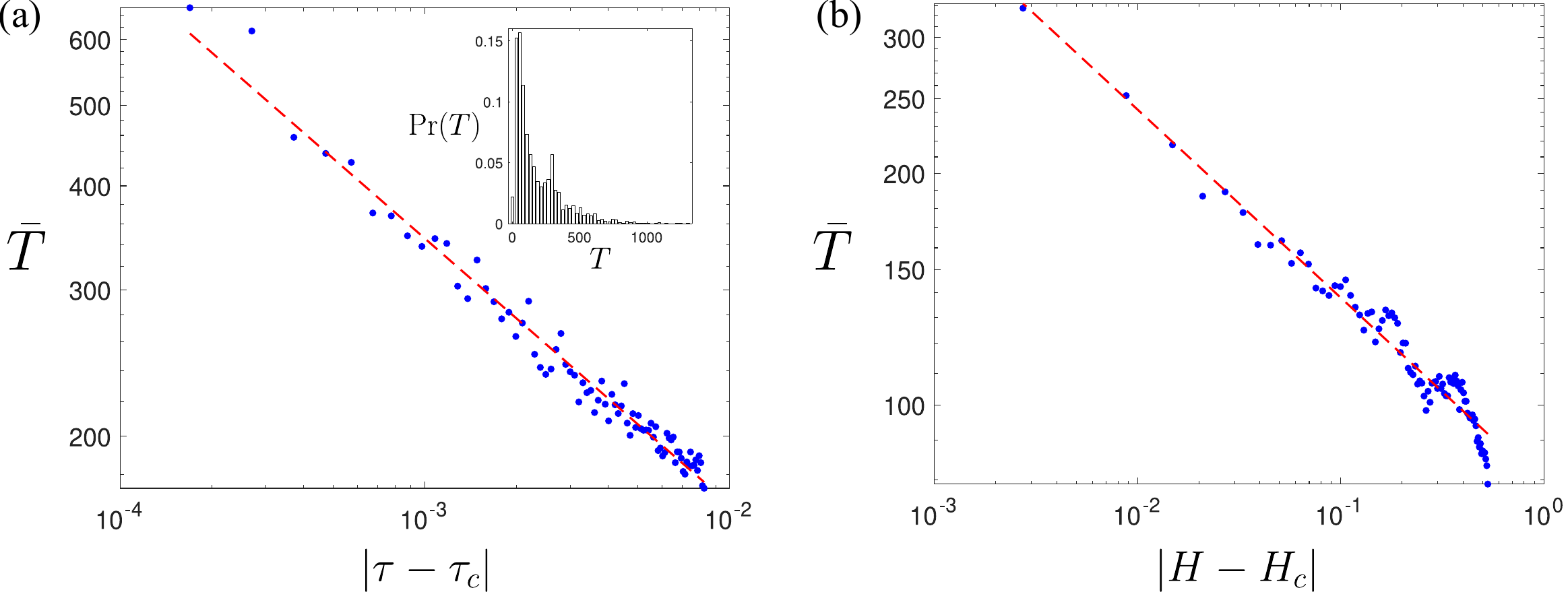}
\caption{Scaling laws for tunneling-like intermittent motion shown in Fig.~\ref{Fig: intermittency}. (a) Scaling behavior near the onset of intermittent dynamics as a function of memory parameter. The critical value of memory parameter for the onset of intermittent behavior is $\tau_c\approx 11.20175$ and the plot shows that the scaling behavior of the mean residence time $\bar{T}$ in each well scales as $\bar{T} \sim |\tau-\tau_c|^{-\gamma}$ with $\gamma\approx 0.32$. The inset shows a typical probability distribution of the residence time $T$ in each well for $\tau=11.21$.  (b) Scaling behavior with changes in height $H$ of the double-well potential for fixed $\tau=11.21$. The critical value of height for the onset of intermittent behavior is $H_c\approx10.03$ and the plot shows that the scaling behavior of the mean residence time $\bar{T}$ follows $\bar{T}\sim |H-H_c|^{-\delta}$ with $\delta\approx 0.24$.}
\label{Fig: scaling}
\end{figure*}

\subsection{Transition to chaos}\label{Sec: Transition to chaos}

%Talk about the transition from stationary state to limit cycles and period doubling route to chaos within a single well. Then ultimately, we get intermittency... I can here include a bifuraction diagram which is consistent with the parameter values elsewhere in the paper. For the bifurcation diagram, it would be good to do this with memory $\tau$, keeping $R=3$ and constant $A$ and $B$.

As the memory parameter $\tau$ is increased keeping other parameters fixed, the initial stationary state of the WPE in the double-well undergoes a series of bifurcations. At the onset of the instability of the stationary states we find limit cycle oscillations. With increasing $\tau$, the limit cycles transition to chaotic motion. The transition to chaos is depicted in Fig.~\ref{Fig: bif} for a typical parameter set. The blue and cyan points represent different but symmetric initial condition of a WPE initialized near the minima of each well. The bifurcation parameter $x_n$ denotes the particle position at times $t_n$ i.e. $x_d(t_n)$ where the times $t_n$ correspond to the Poincaré section given by the constraint $X(t_n)=0$ (i.e., zero velocity of the particle). For aesthetic purposes, we note that both maxima and minima map have been superimposed here, disregarding which specific direction the Poincaré section is traversed.

With increasing memory, we see that the stationary state of the WPE in the minima of the double well becomes unstable via a Hopf bifurcation and then the WPE dynamics takes a period doubling route to chaos. When the WPE dynamics initially become chaotic, the WPE motion is confined to a single well of the double-well potential for $\tau\lesssim 5$. Beyond $\tau\approx 5$, the WPE motion alternates between periodic and chaotic regimes but the WPE motion is no longer confined to a single well and starts to transition between the two wells of the double-well potential in an intermittent fashion. We now explore the nature of this chaotic intermittency between the two wells in detail.

\section{Unpredictable tunneling-like behavior}\label{Sec: tunneling}

A typical intermittent trajectory where the WPE erratically transitions between the two wells of the double-well potential is shown in Fig.~\ref{Fig: intermittency}(a) along with a phase-plane trajectory in Fig.~\ref{Fig: intermittency}(b). Here, the WPE dynamics within each well are chaotic in addition to the chaotic switching between the two wells. This can be seen in the phase-plane where the two chaotic attractors corresponding to each well can be identified clearly, along with links between them that facilitate the inter-well transitions. A passive damped particle in a double-well potential settles down at one of the minima of the well. Conversely, here we find that the interaction of the active particle with its self-generated wave-field, results in intermittent tunneling-like dynamics across the central potential barrier of the double-well forming an analog of quantum tunneling. We note that a similar analog of tunneling was explored by Nachbin \textit{et al.}~\citep{tunnelingnachbin} using their state-of-the-art hydrodynamic pilot-wave model for a walking droplet confined between two cavities. Here, we find a similar tunneling behavior in a simple Lorenz-like model of the WPE system.

The emergence of intermittent tunneling-like behavior shown in Fig.~\ref{Fig: intermittency}(a) can be understood in terms of a crisis phenomena in our Lorenz-like dynamical system. Near the value of $\tau=11.21$ corresponding to Fig.~\ref{Fig: intermittency}(a), we see evidence of a crisis in the bifurcation diagram in Fig.~\ref{Fig: bif}. Below the estimated critical value of $\tau_c \approx 11.20175$, we observe periodic dynamics of the WPE confined in a single well. Then, suddenly, above this critical value, the dynamics of the WPE transitions to chaos. This sudden transition in the bifurcation diagram is characteristic of a symmetry restoring crisis and the intermittent tunneling-like behavior that we observe in Fig.~\ref{Fig: intermittency}(a) is a consequence of crisis-induced intermittency~\citep{Crises}.

To understand the statistics of this intermittent tunneling dynamics, we calculate the time spent by the WPE in each well between transitions i.e. the residence time $T$. The probability density of the residence times $T$ corresponding to the trajectory in Fig.~\ref{Fig: intermittency}(a) is shown in the inset of Fig.~\ref{Fig: scaling}(a). This probability distribution resembles an exponential decay suggesting that very large residence times are unlikely. To understand how the mean residence time $\tilde{T}$, i.e., the time-averaged residence time, varies with the memory parameter near the onset of the intermittency crises, Fig.~\ref{Fig: scaling}(a) shows the mean residence time $\tilde{T}$ as a function of the distance of the memory parameter $\tau$ from the critical memory parameter $\tau_c$. Here, we see a power-law scaling of the form $\bar{T}\sim |\tau-\tau_c|^{-\gamma}$, which is characteristic of crisis-induced intermittency \citep{Crises}. Numerical explorations reveal an approximate value of the scaling exponent $\gamma\approx 0.32$. 
%Frequently, crises befall when the unstable manifold of a saddle fixed point inside the basin of attraction of a particular attractor, hits the basin boundary separating it from another attractor. A value below $1/2$ indicates that the tangency of such unstable manifold is not as it appears in smooth quadratic tangencies of two or higher dimensional Poincaré sections~\citep{Crises}. 
This value is consistent with that of Lorenz-like systems for which there is an unstable time-independent equilibrium solution of the equations (see Sec. IV(b) in Ref.~\citep{Crises} for a more detailed explanation).

To understand the influence of barrier height $H$ of the double-well potential on this intermittent tunneling behavior, we calculate the mean residence time $\tilde{T}$ as we vary the barrier height $H$. The intermittent behavior ceases above the critical value of the barrier height $H_c\approx10.03$. By plotting the mean residence time as a function of distance to this critical value as shown in Fig.~\ref{Fig: scaling}(b), we again obtain a power-law scaling of the form $\bar{T}\sim |H-H_c|^{-\delta}$ with the critical exponent $\delta\approx 0.24$. This suggests that the probability of tunneling between the two wells approaches zero in a power-law manner as the height of the central barrier approaches a critical height (keeping other parameters fixed). This is different to what is typically observed in quantum tunneling. A quantum particle has a non-zero tunneling probability for finite barrier height, while for the WPE we observe that the tunneling like behavior ceases for a finite critical barrier height. However, we note that, it would be possible to retain the intermittent tunneling behavior for increasing barrier height in our WPE system by varying the intrinsic parameters of the WPE i.e. the memory parameter $\tau$ and the wave-amplitude parameter $R$. 

%\subsection{Chaotic scattering}

\begin{figure}
\centering
\includegraphics[width=\columnwidth]{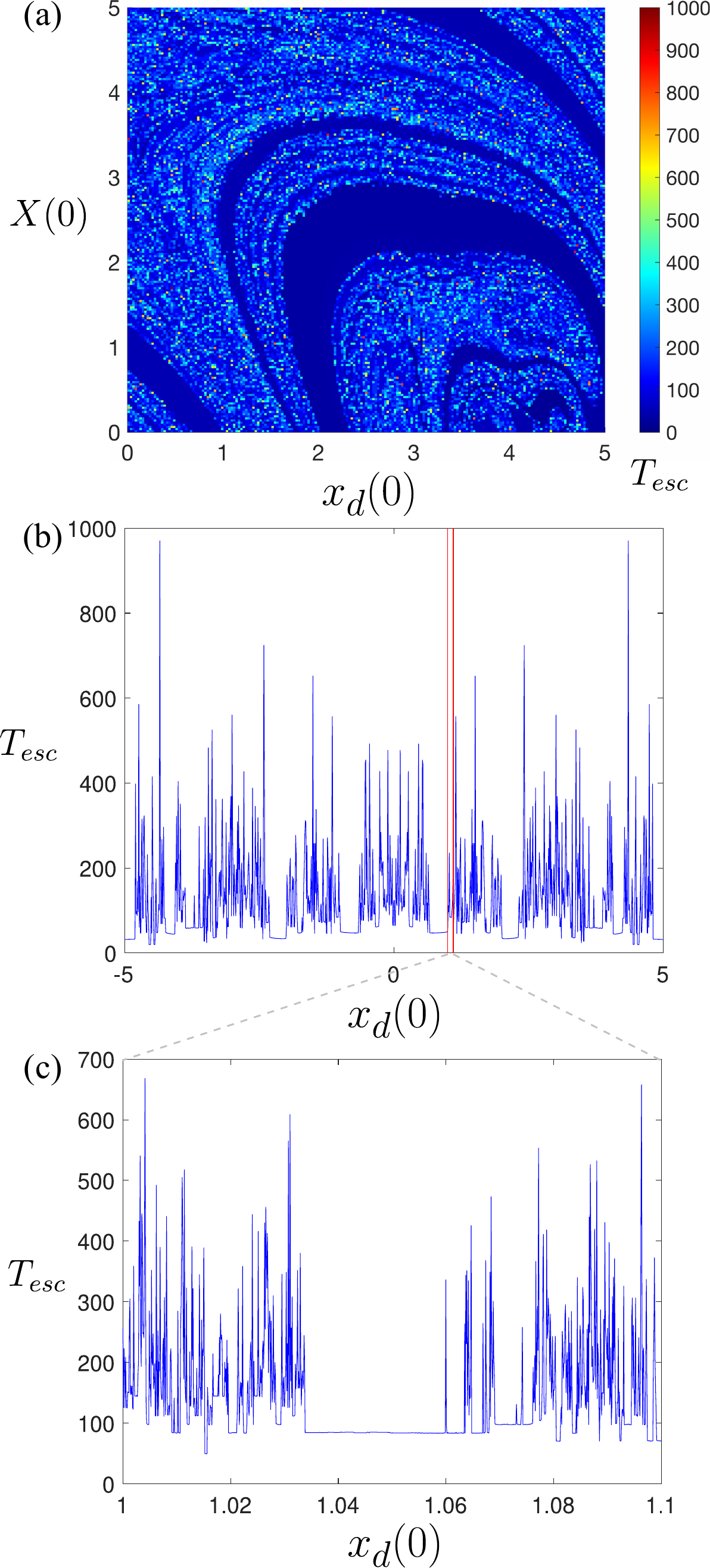}
\caption{Fractal distribution of escape times in the initial-condition space. (a) Colormap showing the fractal distribution of escape times $T_{esc}$ in the initial-condition plane $(x_d(0),X(0))$. (b) One-dimensional distribution of $T_{esc}$ as a function of $x_d(0)$ (for fixed $X(0)=0$) and its refined plot in panel (c) showing the fractal-like structure. Parameter values are $R=1.2$, $\tau=11.21$, $B=2$ and $A=B^2/(4H)$ with $H=10$.}
\label{Fig: fractal}
\end{figure}

\begin{figure}
\centering
\includegraphics[width=0.9\columnwidth]{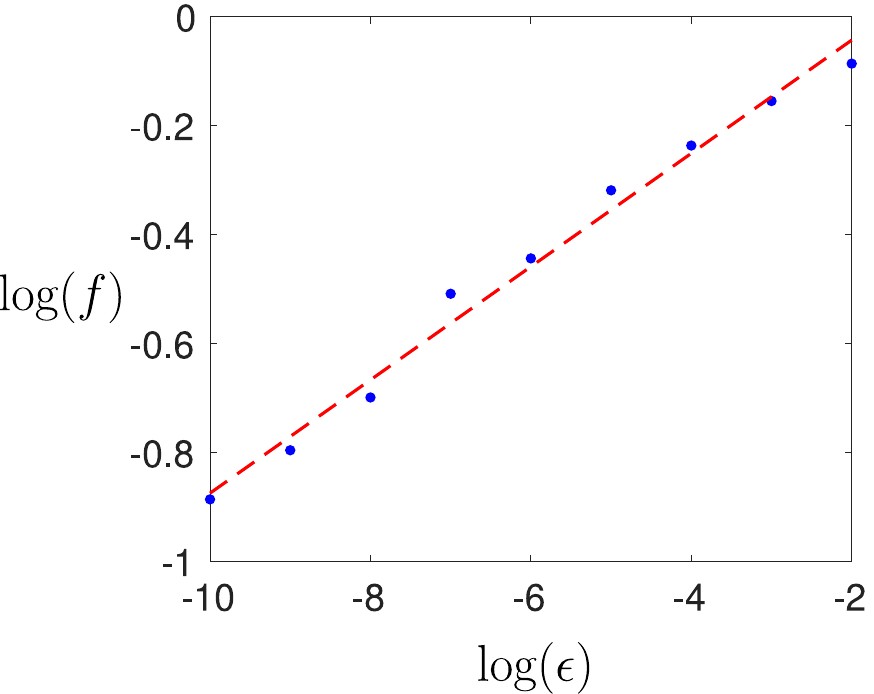}
\caption{Unpredictability of intermittent tunneling behavior. A plot of the logarithm of fraction $f$ of events that are uncertain, against the distance $\epsilon$ between the points $x_d(0)$ and $x_d(0)+\epsilon$. Two points are considered uncertain if $|T_{esc}(x_d(0))-T_{esc}(x_d(0)+\epsilon)|>h$, with $h=1.0$. The uncertainty exponent $\nu$ can be computed from the relation $f(\epsilon) \propto \epsilon^\nu$, and gives a value of $\nu\approx0.10$.}
\label{Fig: unc}
\end{figure}

\begin{figure*}[t]
\centering
\includegraphics[width=1.7\columnwidth]{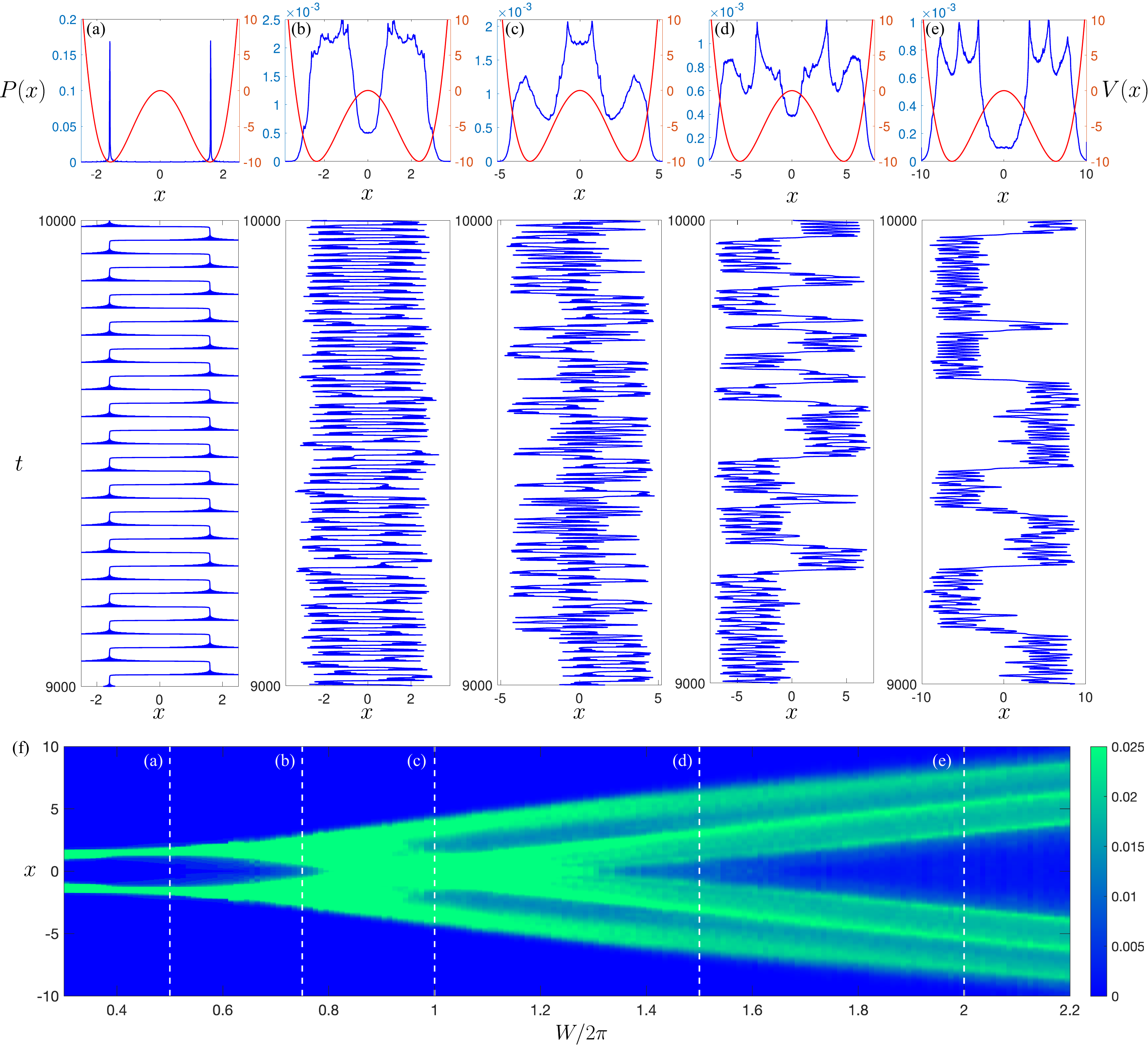}
\caption{Emergence of wave-like statistics from underlying chaos. Probability distribution $P(x)$ of finding the particle in the double well potential (top panel) and its space-time trajectory (bottom panel) as a function of the width $W$ of the double well. We fix $R=3$, $\tau=100$ and height $H=B^2/4A=10$. Panels (a)-(e) correspond to $W=0.5 \times 2\pi$, $W=0.75 \times 2\pi$, $W=1 \times 2\pi$, $W=1.5 \times 2\pi$ and $W=2 \times 2\pi$, where $2\pi$ is the wavelength of the particle-generated waves. Here $A=64H/W^4$ and $B=16H/W^2$. (f) Probability distribution (colormap) $P(x)$ as a function of the rescaled width, $W/2\pi$, of the double-well potential. The vertical dashed line correspond to the probability distributions in panels (a)-(e). See supplemental videos S2-S5 for movies of WPE dynamics corresponding to panels (b)-(e) respectively.}
\label{Fig: wavelike stat chaos}
\end{figure*}

We now investigate the sensitivity of the intermittent tunneling dynamics to the initial conditions of the system. This is done by calculating the escape time $T_{esc}$ required by the particle to cross the central maximum at $x=0$ for the first time. By fixing the initial values of the variables $Y(0)=Z(0)=0$ and plotting the value of escape time in the initial condition plane $(x_d(0),X(0))$, we obtain the distribution shown in Fig.~\ref{Fig: fractal}. We observe a fractal distribution of escape times where small escape time coexist with anomalously large escape times. The fractal nature of the distribution is further evident by plotting a one-dimensional distribution along $X(0)=0$, as depicted in Figs.~\ref{Fig: fractal}(b) and (c), resembling a Cantor-like set. All this suggests that for two nearby initial conditions, one may get very different escape times making the intermittent tunneling transitions between wells uncertain. This can be quantified by calculating the uncertainty exponent, allowing us to approximate the fractal dimension of this one-dimensional Cantor-like set~\citep{GREBOGI1983415}. The uncertainty exponent in our example measures the probability that two points with different initial positions in the phase space $x_d(0)$ and $x_d(0) +\epsilon$, have a given difference in their value of their escape times $|T_{esc}(x_d(0))-T_{esc}(x_d(0)+\epsilon)|$. Two points that exceed a certain small threshold $h$, are said to be uncertain. A scaling law can then be obtained by plotting the fraction of points that are uncertain for a range of distances in the initial positions of the randomly chosen pairs of points. The exponent of the power-law is called the uncertainty exponent, and measures how much the escape times fluctuate in the $x_d(0)$ initial position space. The scaling behaviour is shown in Fig.~\ref{Fig: unc} from which we calculated the uncertainty exponent as $\nu\approx0.10$, giving us a fractal dimension of $D_F=1-\nu\approx0.90$. This value suggests great unpredictability in the first passage times.

\section{Wave-like statistics}\label{Sec: wavelike stats}

It is clear from the previous analyses that there exists an intimate connection between chaotic dynamics and the statistics of tunneling behavior. Emergence of wave-like statistics from underlying chaotic dynamics has been observed in experiments and simulations with walking droplets confined in a small domain whose extent is a few times the wavelength of droplet-generated waves~\citep{PhysRevE.88.011001,Cristea,durey_milewski_wang_2020,Giletcircularcavity,Giletcircularcavity2,Tambascoorbit,durey2018}. Wave-like statistics emerge in these confined systems when (i) the decay time (i.e. memory) of the waves generated by the droplet is longer than the time required for the droplet to cross its domain so that the droplet continually interacts with its self-excited wave field and (ii) the droplet's dynamics are chaotic and it switches intermittently between different unstable periodic orbits~\citep{PhysRevE.88.011001,Saenz2017}. In this section we explore the emergence of wave-like statistics in the chaotic regime of our simple Lorenz-like model of a WPE in a double-well potential.% We investigate the statistics emerging from the underlying chaotic dynamics, which are shown in Fig.~\ref{Fig: wavelike stat chaos}. 

\begin{figure*}
\centering
\includegraphics[width=2.0\columnwidth]{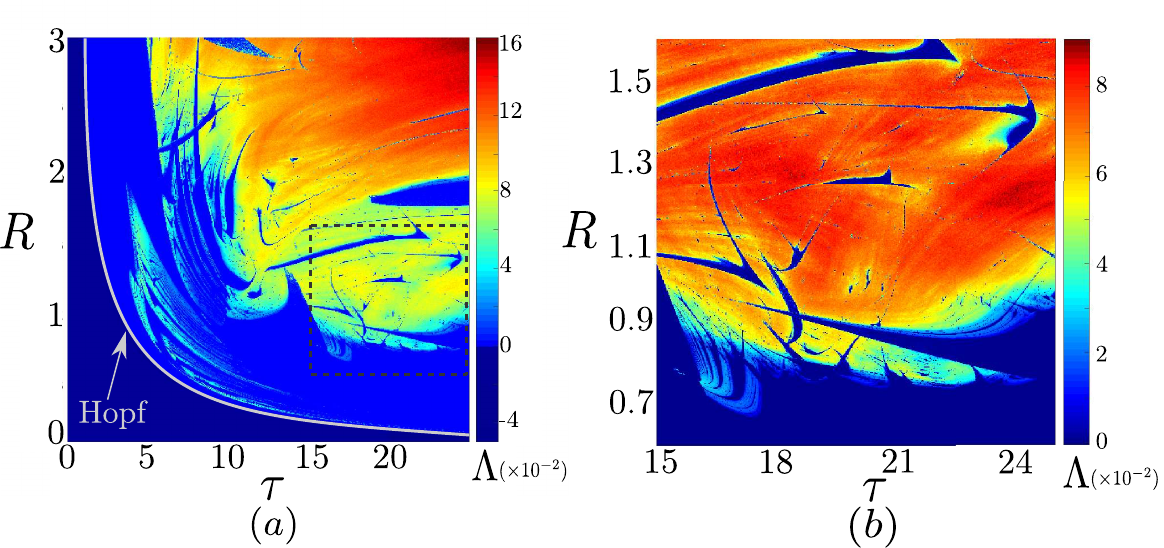}
\caption{Chaotic regions in the $(\tau,R)$ parameter-space for $B=2$, $H=10$ and $A=B^2/(4H)$. The largest Lyapunov exponent $\Lambda$ is shown in the parameter space. (a) A region where stationary (dark blue) and, limit cycle periodic and chaotic motion (light blue to red) are present. (b) An inset of the previous chaotic parameter set with slightly modified color scheme (small values of $\Lambda$ are depicted in dark blue, representing periodic complex limit cycle oscillations, and a more clear chaotic sea is shown as the colors become more light. These scheme allows to distinguish clearly chaotic and periodic parameter space regions.}
\label{Fig: PS}
\end{figure*}

When the amplitude of the waves is not too small, at large values of the memory parameter $\tau$, chaotic motion of the WPE tends to dominate with increased robustness of the intermittent dynamics between the two wells of the double-well potential (see Fig.~\ref{Fig: PS}). Figure~\ref{Fig: wavelike stat chaos}(a)-(e) shows the probability distribution of the particle position (top panel) along with the particle trajectory (bottom panel) as the width of the double-well potential is increased, at a fixed high value of memory parameter of $\tau=100$. For small widths of the double-well potential, $W=2\pi\times 1/2$, i.e. half the wavelength of the particle-generated waves, we observe a periodic trajectory where the WPE spends a fixed time at the minima of the well and then abruptly transitions to the other minima (see Fig.~\ref{Fig: wavelike stat chaos} (a)). Increasing the width of the double-well potential to $0.75$ of the WPE's wavelength results in chaotic motion. However, the probability distribution reveals that the WPE still spends a large fraction of its time near the minima of the two wells and less time near the maxima of the well, as shown in Fig.~\ref{Fig: wavelike stat chaos}(b). With increasing width of the double-well potential, we see an increase in the number of peaks in the wave-like probability distribution as it can be seen in Fig.~\ref{Fig: wavelike stat chaos}(c)-(e). This behavior qualitatively resembles the probability distribution for a quantum particle in a double-well potential. With increasing energy level, the number of peaks in the probability distribution of the quantum particle also increase similar to Fig.~\ref{Fig: wavelike stat chaos}. However, we note that for our WPE system, these different wave-like probability distributions are obtained by varying the width of the double-well potential~(see Fig.~\ref{Fig: wavelike stat chaos}(f)), while for a quantum particle, these probability distributions are realized for a fixed double-well potential and different energy levels.

To fully capture the nature of chaotic dynamics in the $(\tau,R)$ parameter space corresponding to the particle's internal wave-like properties when $A$ and $B$ are fixed, we have calculated the largest Lyapunov exponent (LLE) in this parameter space and plotted it for typical parameter values as shown in Fig.~\ref{Fig: PS}. In this plot, the non-chaotic regime is represented by dark blue regions where LLE takes a value of zero whereas a not too small positive value of LLE indicate chaotic behavior with increasing sensitivity to initial conditions for large values of LLE. Chaos tends to dominate in the high memory limit, with periodic islands being less frequent. We also observe a fractal distribution of periodic islands in this parameter space, where chaotic and non-chaotic regions are intricately interwoven. Self-similar shrimp-like fractal structures of periodic orbits embedded in chaotic sea are evident. These structures are typical in nonlinear discrete dynamical systems~\citep{Shrimps1}. They are related to how periodic orbits organize in the parameter space and  have been observed to arise in Lorenz-like systems as well~\citep{Shrimps2}. These features highlight the dynamical complexity arising even in our simple finite-dimensional model of a WPE system. 
%\begin{figure}
%\centering
%\includegraphics[width=\columnwidth]{figps.eps}
%\caption{Different dynamical behaviours in the $(B,\tau)$ parameter-space for $R=3.0$ and $B^2/4A=10$. Blue indicates periodic/stationary particle trajectories confined to a single well, Red indicates periodic trajectories that crosses the well, yellow indicates chaotic trajectories inside a single well and purple indicates chaotic trajectories that go through both wells.}
%\label{Fig: PS3}
%\end{figure}

We can further simplify algebraically our dynamical system in the high memory regime by taking the limit $\tau \xrightarrow{} \infty$. This simplifications yields the reduced dynamical system
\begin{equation}
    \label{lorenz inf}
    \begin{split}
    \dot{x}_d&=X, \\
    \dot{X}&=Y-X-A x^3_d + B x_d, \\
    \dot{Y}&=XZ, \\
    \dot{Z}&=R-XY.
  \end{split}
\end{equation}
In this limit, the self-generated waves of the particle do not decay in time. Nevertheless, we also do not get a divergence of the particle's position, since the self-generated waves interfere both constructively and destructively, keeping the particle's motion bounded~\citep{Valani_inf_mem}. In this infinite-memory regime, the wave-like statistical behavior observed in Fig.~\ref{Fig: wavelike stat chaos} still persists with similar probability distributions. 
%To understand the variation in the system dynamics with the wave-amplitude parameter $R$, we fixed the external double-well potential and plot the bifurcation diagram as a function of $R$ as shown in Fig.~\ref{Fig: bif inf}. In this bifurcation diagram we mainly observe chaotic behavior with small regions of periodicity. 
Indeed, despite the increased chaoticity in our system, we find that the intertwined regions of periodicity and chaos still persist in the $(R,\tau$) parameter plane as $\tau\xrightarrow{}\infty$.%, as far as we have been able to compute.

\section{Discussion and Conclusion}\label{Sec: Discussion and Conclusion}

We have explored the complex dynamical and quantum-like features arising in the system of a one-dimensional WPE in a symmetric double-well potential. A Lorenz-like system models the dynamical behaviors which we analyzed in detail. 

At low values of the memory parameter $\tau$ and the wave amplitude parameter $R$, we observe stable stationary states of the WPE located at the two symmetric minima of the double-well potential. With increasing memory, the stationary states become unstable via Hopf bifurcation and limit cycle oscillations appear. The amplitude of these highly nonlinear oscillations increases in a quantized manner as the width of the potential is increased. Importantly, these orbits are asymptotically attractive, in contrast with typical orbits of linear and nonlinear conservative dynamical systems. This phenomenon constitutes an analog of quantized eigenstates.

Furthermore, we have demonstrated an alternate analog of quantized eigenstates, where we have found four distinct limit cycle states existing at fixed parameter values. The system exhibits multistability with intertwined basins. Surprisingly, the mean Lyapunov energies for these states display energy splitting with two limit cycle states showing closely spaced energy levels. This feature is characteristic of a quantum particle in a double-well potential. 

To form a more robust analog of quantized eigenstates, one would ideally need a countably infinite number of coexisting limit cycle attractors to achieve a discrete and infinite energy spectrum. In fact, this notion is possible in nonlinear dynamical systems and it is known as megastability~\citep{Sprott2017}. Classical systems have Hamiltonian structure that results in an uncountably infinite number of periodic orbits with a continuous energy spectra. Conversely, quantum systems have a countably infinite number of discrete eigenstates and energy levels. The notion of megastability in nonlinear dynamical systems provides a dynamical mechanism to obtain countably infinite set of states. The existence of megastability in dynamical systems of WPE, and in general delay systems, is considered elsewhere~\citep{Lopezvalanimegastablity}. 
% \begin{figure}[b]
% \centering
% \includegraphics[width=\columnwidth]{Fig11.pdf}
% \caption{Bifurcation diagram of a WPE in a double-well potential as a function of the wave amplitude parameter $R$ in the infinite-memory regime. The values $x_n$, particle position when the instantaneous velocity is zero, are plotted as a function of $R$ for two different but symmetric initial conditions (blue and cyan). The external double-well potential is fixed with parameters $B=2$, $H=10$ and $A=B^2/4H$.}
% \label{Fig: bif inf}
% \end{figure}

With increasing memory parameter $\tau$, we found that the dynamics of the WPE undergo a period doubling transition to chaos within a single well. Beyond this regime, we observed intermittent trajectories of the WPE that erratically switch between the two wells of the double-well. We rationalized this intermittent chaotic dynamics in terms of crisis-induced intermittency and showed a power law scaling near the onset of intermittency. We further demonstrated unpredictability in the switching dynamics between the two wells by calculating the escape times of the WPE, which showed a fractal structure. This unpredictable intermittent transition of the WPE between the two wells constitutes an analog of quantum tunneling. The intermittent tunneling behavior resulted in emergence of wave-like statistics where the particle's probability distribution displayed wave-like features with increasing width of the potential. These probability distributions were qualitatively similar to that of a quantum particle in a double-well potential. We also observed rich dynamical features in the parameters space where shrimp-like islands of periodicity exist in the chaotic sea. 

We note that the theoretical model explored in this manuscript is different from the experimental setup of Eddi \textit{et al.}~\cite{Eddi2009} for demonstrating an analog of tunneling using walking droplets. In their experiments, the ``potential barrier" is implemented via a submerged physical barrier which interacts with the walking droplet through its self-generated wave field. Conversely, in our system, the external double well potential is only acting on the particle and not influencing the particle-generated waves. In this respect, an experimental realization of the dynamics presented here might be achieved using an experimental setup similar to that of Perrard \textit{et al.}~\cite{Perrard2014a} where they used ferrofluids and external magnetic fields to impart an external harmonic potential directly to the droplet. To consider a set-up analogous to  Eddi \textit{et al.}~\cite{Eddi2009} in our simple theoretical model, one can implement the ``submerged barrier" by abruptly changing $R$ and $\tau$ values to below the walking threshold in the shallow region of Eddi~\textit{et al.}~\cite{Eddi2009}.

The present exploration shows that rich dynamical and quantum-like features such as multistable quantized states, intermittent tunneling and wave-like statistics can be observed in a simple Lorenz-like model of a WPE in a double-well potential. It would be fruitful to explore this double-well potential setup in experiments with walking and superwalking droplets to see if some of these behaviors can be reproduced, and to understand their robustness as the memory of the system enhances.

%The wave-like statistics emerged from intermittent chaotic dynamics
%Hence, in general if one considering all the unstable limit cycles of the chaotic system and assigns different weights to them which represent the time spent by a phase-space trajectory on those unstable limit cycles, then one maybe able to reverse engineering and obtain a desired probability distribution by appropriately assigning different weights to the unstable limit cycles.

%%
 \bibliographystyle{elsarticle-num} 
 \bibliography{cas-refs}




\biboptions{sort&compress}
% \end{thebibliography}
\end{document}